\documentclass[aps, prl, superscriptaddress, twocolumn,amsmath,amssymb,floatfix]{revtex4-1}

\usepackage{hyperref}
\usepackage{graphicx}
\usepackage{dcolumn}
\usepackage{xcolor}
\usepackage{commath}
\usepackage{mathtools}
\usepackage[normalem]{ulem}

\renewcommand{\Re}{\operatorname{Re}}
\renewcommand{\Im}{\operatorname{Im}}
    
\begin{document}

\title{Non-Hermitian dynamics and nonreciprocity of optically coupled nanoparticles}

\author{Manuel Reisenbauer}
\thanks{These authors contributed equally.}
\affiliation{Vienna Center for Quantum Science and Technology (VCQ), Faculty of Physics, University of Vienna, Boltzmanngasse 5, A-1090 Vienna, Austria}

\author{Henning Rudolph}
\thanks{These authors contributed equally.}
\affiliation{Faculty of Physics, University of Duisburg-Essen, Lotharstra\ss e 1, 47048 Duisburg, Germany}

\author{Livia Egyed}
\thanks{These authors contributed equally.}
\affiliation{Vienna Center for Quantum Science and Technology (VCQ), Faculty of Physics, University of Vienna, Boltzmanngasse 5, A-1090 Vienna, Austria}

\author{Klaus Hornberger}
\affiliation{Faculty of Physics, University of Duisburg-Essen, Lotharstra\ss e 1, 47048 Duisburg, Germany}

\author{Anton V. Zasedatelev}
\affiliation{Vienna Center for Quantum Science and Technology (VCQ), Faculty of Physics, University of Vienna, Boltzmanngasse 5, A-1090 Vienna, Austria}

\author{Murad Abuzarli}
\affiliation{Vienna Center for Quantum Science and Technology (VCQ), Faculty of Physics, University of Vienna, Boltzmanngasse 5, A-1090 Vienna, Austria}

\author{Benjamin A. Stickler}
\affiliation{Institute for Complex Quantum Systems, Ulm University, Albert-Einstein-Allee 11, D-89069 Ulm, Germany}

\author{Uro\v{s} Deli\'{c}}
\email{uros.delic@univie.ac.at}
\affiliation{Vienna Center for Quantum Science and Technology (VCQ), Faculty of Physics, University of Vienna, Boltzmanngasse 5, A-1090 Vienna, Austria}

\begin{abstract}

Non-Hermitian dynamics, as observed in photonic, atomic, electrical, and optomechanical platforms, holds great potential for sensing applications and signal processing. Recently, fully tunable nonreciprocal optical interaction has been demonstrated between levitated nanoparticles. Here, we use this tunability to investigate the collective non-Hermitian dynamics of two nonreciprocally and nonlinearly interacting nanoparticles. We observe parity-time symmetry breaking and, for sufficiently strong coupling, a collective mechanical lasing transition, where the particles move along stable limit cycles. This work opens up a research avenue of nonequilibrium multi-particle collective effects, tailored by the dynamic control of individual sites in a tweezer array.

\end{abstract}

\maketitle

A plethora of physical phenomena are well described by Hermitian dynamics, such as the dynamics of closed quantum systems, Landau-type phase transitions, or transport along oscillator chains. However, the growth of complexity in quantum many-body systems, the occurrence of chiral transport properties, and the emergence of experiments that strongly couple to the environment require models that go beyond Hermitian descriptions. A particularly interesting example of this so-called non-Hermitian dynamics are nonreciprocal interactions, which seemingly break Newton's third law of action equals reaction. Besides being often encountered in biological systems \cite{Fruchart2021}, nonreciprocity -- in its broadest sense -- has found various applications in optics and photonics \cite{Feng2017, Xiao2020, Weidemann2020, Wang2021, Guo2009, Zhen2015, Miri2019}, ultracold atoms \cite{Li2019, Gou2020, Takasu2020, Oeztuerk2021, Ferri2021, Liang2022,Murch2023}, electrical circuits \cite{Helbig2020, Zou2021}, and metamaterials \cite{Brandenbourger2019, Ghatak2020, Chen2021, Wang2022}. The size and sensitivity to environmental perturbations make nonreciprocally interacting arrays of mechanical objects an ideal ground for realizing unidirectional or topological transport \cite{Metelmann2015, Fang2017, SanavioXuereb}, enhanced sensing due to nonreciprocity \cite{MetelmannQLA, LauClerk, McDonald2020}, and topological states \cite{PeanoMarquardt, Rosenthal2018, Wanjura2020}. So far, there have been several experimental demonstrations of nonreciprocal or non-Hermitian dynamics in various platforms in the classical regime \cite{MathewEwold, XuHarris2016, Patil2022, RenPainter, Zhang2022, YoussefiKippi, DosterWeig, Zheludev2023}.

Optically levitated nanoparticles have become a well-established system for quantum physics with translational and rotational degrees of freedom \cite{GonzalezBallestero2023, SticklerReview}. Recently, there has been a surge of experiments that extend trapping and control from single particles to particle arrays in a variety of geometries \cite{Rieser2022, Vijayan2022, Yan2023, LiskaColdDamping, vijayan2023cavitymediated, BykovOptica}, thus demonstrating that this platform is highly versatile and scalable. In one of those experiments, we have demonstrated direct, nonreciprocal, and nonlinear light-induced dipole-dipole interactions between particles in a tweezer array \cite{Rieser2022}. Such optically interacting particle arrays offer several benefits for investigations of (quantum) non-Hermitian physics. For example, single-site readout enables the full reconstruction of the collective degrees of freedom \cite{PennyPRR, LiskaColdDamping, BykovOptica}. At the same time, the optically induced forces allow for a wide tuning range from reciprocal to unidirectional to anti-reciprocal interactions \cite{Rudolph2023}. Altogether, this system enables studies in previously unexplored interaction regimes with an unprecedented level of control.

In this work, we report on the experimental investigation of non-Hermitian dynamics stemming from the anti-reciprocal interaction between the motion of two trapped silica nanoparticles. We observe two exceptional points (EP) that define a region where the collective motion is in the parity and time-reversal ($\mathcal{PT}$) symmetry-broken phase. In this phase, the interaction leads to correlated particle motion, which we confirm by measuring a constant phase delay between the oscillators. The system further exhibits a Hopf bifurcation into the mechanical lasing phase, where the interaction-induced amplification dominates over the intrinsic damping such that the motion becomes nonlinear.  

\begin{figure*}[ht!]
    \centering
    \includegraphics[width=\linewidth]{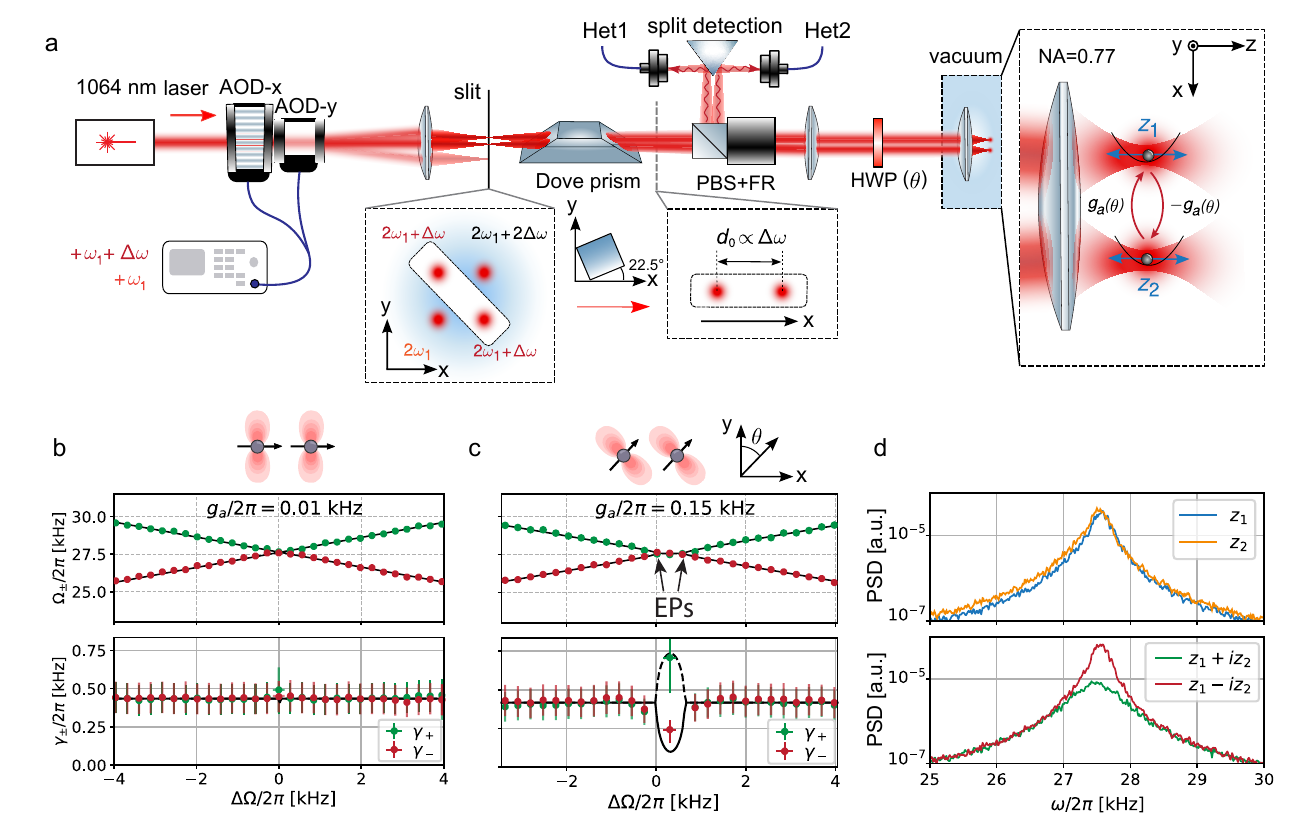}
    \caption{\textbf{Experimental setup.} \textbf{a.} Each of the two orthogonal acousto-optical deflectors (AOD-x/y) is driven by the same two radio-frequency (RF) tones at frequencies $\omega_1$ and $\omega_1+\Delta\omega$, which creates $2\times 2$ laser beams. We use a slit to select the two beams that have equal optical frequencies, while the distance between the beams can be tuned by changing the RF tone frequency difference $\Delta \omega$. The Dove prism rotates the optical plane to place the beams into the plane of the optical table. The beams are subsequently focused in the vacuum chamber to form two traps at a distance $d_0$. The light back-scattered by the particles is reflected with a Faraday rotator (FR) and a polarizing beamsplitter (PBS) and sent to two independent heterodyne detectors that monitor the particle motion. Inset: Two particles are trapped and interact anti-reciprocally with the coupling rate $\pm g_a$ tuned by the polarization angle $\theta$, which we set with a half-wave plate (HWP) in front of the vacuum chamber. \textbf{b.} For polarization along the $x$-axis ($\theta\approx \pi/2$), the interaction between the particles is weak such that the two modes cross (green and red points: measured eigenfrequencies and damping rates, black lines: fits). \textbf{c.} For $\theta\neq \pi/2$ we observe degenerate eigenfrequencies between the two exceptional points (EP) (top) and nondegenerate damping rates that are split by $4g_a$ (bottom). Black lines are theory functions based on the measured coupling rate. \textbf{d.} At the maximum splitting of the damping rates we can reconstruct the power spectral densities (PSD) of the eigenmodes of the particles' positions as $z_1\pm i z_2$ (bottom) from the detected positions $z_{1,2}$ (top).}
  
    \label{fig1:setup}
\end{figure*}

\paragraph{\textbf{Experimental setup.}} We use a pair of orthogonal acousto-optical deflectors (AODs), both driven by two radio-frequency (RF) tones at frequencies $\omega_1$ and $\omega_1+\Delta\omega$, to create $2\times 2$ laser beams from a laser source at a wavelength of $\lambda = 1064~\text{nm}$ (Fig.~\ref{fig1:setup}a). We use a narrow slit (width: $2~\text{mm}$) to select the laser beams on the diagonal as they have equal laser frequencies, which is required to control the optical interaction. A Dove prism tilted at an angle of $22.5^{\circ}$ rotates the optical plane by $45^{\circ}$, thus placing the laser beams in the plane of the optical table. A high numerical aperture lens ($\text{NA}=0.77$) inside the vacuum chamber focuses the laser beams to two foci at a relative distance $d_0$, in which we trap two silica nanoparticles of approximately equal sizes (nominal radius $r=(105\pm 2)~\text{nm}$ \cite{SI}). We tune the particle distance with the frequency difference of the RF tones as $d_0\propto \Delta\omega$ on both AODs, which maintains equal laser frequencies of the two tweezers \cite{SI}. The optical phases at the traps $\phi_{1,2}$ are controlled by the phases of the RF tones. For optical powers of $P_{1,2}\approx 0.3~\text{W}$ the resulting oscillation frequencies of the center-of-mass (CoM) motion along the tweezer axis ($z$-axis) are $\Omega_{1,2} \propto \sqrt{P_{1,2}}\approx 2\pi\times 27.5~\text{kHz}$. To sweep the mechanical frequency detuning $\Delta \Omega = \Omega_2 - \Omega_1$, we scan powers $P_{1,2}$ symmetrically such that the total power is conserved. We collect the light backscattered from the particles with separate fiber-based confocal microscopes, which allows us to independently detect the particles' motion with balanced heterodyne detections \cite{SI}. To suppress electrostatic coupling we place the particles at a large distance of $d_0\sim 18.4~\mu\text{m}$ and discharge them. The intrinsic damping rate, given by the gas pressure, is kept constant at $\gamma/2\pi = (0.46 \pm 0.02)~\text{kHz}$ throughout the measurement.  

\begin{figure*}[ht!]
    \centering
    \includegraphics[width=0.9\linewidth]{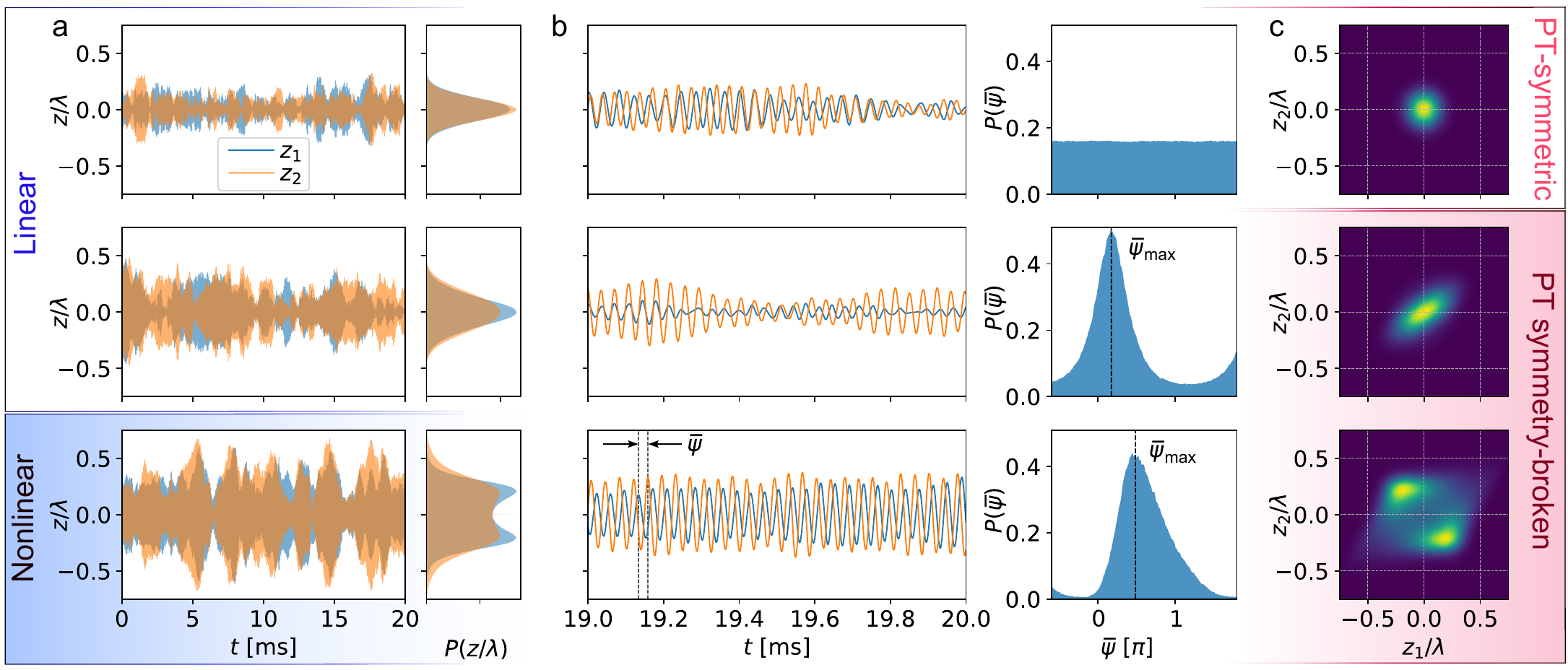}
    \caption{\textbf{Collective statistics of uncoupled (top row), linear (middle) and nonlinear (bottom) particles' motion $\mathbf{z_1}$ (blue) and $\mathbf{z_2}$ (orange).} In the middle and bottom rows the system is in the $\mathcal{PT}$ symmetry-broken phase. \textbf{a.} The histograms of the particles' motion show an increasing variance (top to middle) and eventually the transition from linear into nonlinear motion (bottom). \textbf{b.} Uncoupled particles move independently, which is confirmed by the uniform distribution of the phase delay $\bar{\psi}$ between the oscillators. On the other hand, in the $\mathcal{PT}$ symmetry-broken phase the histograms show a preferred phase $\bar{\psi}_{\text{max}}$ as $z_1$ and $z_2$ are strongly correlated. The black lines mark the most probable phase delay $\bar{\psi}_{\text{max}}$. \textbf{c.} The joint $z_1$-$z_2$ distributions show the transition from a thermal motion (top) to a correlated motion (middle) to a limit cycle (bottom).}
  
    \label{fig2}
\end{figure*}

\paragraph{\textbf{Non-Hermitian Hamiltonian.}} We model the particles' $z$-motion in the frame rotating with the mean oscillation frequency $\Omega_0 = (\Omega_1 + \Omega_2) / 2$. The linearized equations of motion, averaged over one oscillation period $2\pi/\Omega_0$, are:
\begin{equation}
\frac{d}{dt} 
    \begin{pmatrix}
     a_{1}  \\
     a_{2}   \\ 
    \end{pmatrix} = -i H_{\text{NH}} 
    \begin{pmatrix}
     a_{1}  \\
     a_{2}   \\ 
    \end{pmatrix} + \sqrt{ \gamma n_T}     
    \begin{pmatrix}
     \xi _1 \\
     \xi _2  \\ 
    \end{pmatrix},
    \label{eq:equmotion}
\end{equation}
where $a_{j}=  \left( z_j + i p_j/m \Omega_j \right) \exp(i \Omega_0 t)/2z_{\text{zpf},j}$ is the complex amplitude of particle $j$ with position $z_j$, momentum $p_j$, zero point fluctuation $z_{\text{zpf},j}=\sqrt{\hbar/2m\Omega_j}$ and mass $m$. Gas damping and diffusion enters via the gas damping rate $\gamma$, the thermal occupation $n_T$ in equilibrium with the environment and the complex white noises $\xi _{1,2}$ with correlations $\langle \xi ^*_{j'}(t') \xi_{j}(t)\rangle = \delta_{jj'} \delta(t-t')$ and $\langle \xi_{j'}(t') \xi_{j}(t) \rangle = 0$. The effective non-Hermitian Hamiltonian realized by the optical interaction is \cite{Rudolph2023}:
\begin{equation}
    H_{\text{NH}} = \begin{pmatrix}
     - \frac{\Delta \Omega}{2} + g_r + g_a - i\frac{\gamma}{2} &        -(g_r + g_a)\\
     -(g_r - g_a)         & \frac{\Delta \Omega}{2} + g_r - g_a - i \frac{\gamma}{2}\\ 
    \end{pmatrix},
    \label{eq:hamiltonian}
\end{equation}
where $g_r = G \cos{(k d_0)} \cos{(\Delta \phi)} / k d_0$ and $g_a = G \sin{(k d_0)} \sin{(\Delta \phi)} / k d_0$, with $k=2\pi/\lambda$, describe the reciprocal (conservative) and the anti-reciprocal (non-conservative) coupling rate, respectively \cite{Rieser2022, Rudolph2023}. The nonreciprocity of the effective interaction arises from the interference of the tweezer and the light scattered off the particles, which effectively carries away momentum. The interference and thus the coupling rates can be tuned by the optical phase difference $\Delta \phi=\phi_2-\phi_1 $ and the distance $d_0$ between the particles. In our work, we set $d_0$ and $\Delta \phi$ such that $g_r$ is negligible, while $g_a$ takes its maximum value for a given distance. Its magnitude is then determined by the coupling constant $G \propto \sqrt{P_1 P_2} \cos^2(\theta)$ as a function of the trapping powers $P_{1,2}$ and the laser polarization angle $\theta$ (Fig. \ref{fig1:setup}b,c), which we modify with a half-wave plate (HWP) in front of the vacuum chamber \cite{SI}. The eigenvalues of the effective Hamiltonian $H_{\text{NH}}$ are in general complex and are given by $\lambda_{\pm} = - i \gamma /2 \pm \sqrt{\Delta \Omega ^2 - 4 g_a \Delta \Omega}/2$, such that the frequencies and damping rates of the eigenmodes are given by the real and imaginary parts $\Omega_{ \pm } = \Omega_0+\Re( \lambda_{ \pm })$ and $\gamma_{\pm} = - 2\Im( \lambda_{ \pm })$, respectively. The eigenvectors coalesce at the detunings of $\Delta \Omega_{\text{EP}1,2} = 2 g_a \mp 2 g_a$, which define the exceptional points (EP). For $\Delta\Omega$ between the EPs, the frequencies are degenerate and the damping rates become non-degenerate, resulting in the so-called "normal mode attraction" (Fig.~\ref{fig1:setup}c).  The maximum splitting of the damping rates $\gamma_{\pm}$ is achieved for $\Delta\Omega =2g_a$ and is equal to $4g_a$, where the complex eigenmodes of the system can be reconstructed as $a_{\pm}=a_1\pm i a_2$. Note that the sign in the corresponding eigenmodes of the motion $z_{\pm} = z_1\mp i z_2$ is flipped due to the definition of $a_{1,2}$. Therefore, the different damping rates result in the suppression of the eigenmode $a_-$ ($z_-$), while $a_+$ ($z_+$) is amplified (Fig.~\ref{fig1:setup}d). Note that the Hamiltonian $H_{\text{NH}}$ is in general $\mathcal{PT}$-symmetric; however, the $\mathcal{PT}$ symmetry is broken in the region between the EPs as the eigenmodes have different damping rates. 

\paragraph{\textbf{Nonlinear anti-reciprocal interactions.}} Once the effective damping rate $\gamma_-$ becomes negative, the linear theory breaks down. An analytical model that includes the full nonlinear dynamics -- but no thermal fluctuations -- yields a system of differential equations for the amplitude of the collective motional state $ A = 2k\sqrt{\hbar/m\Omega_0}  \sqrt{|a_2|^2 + |a_1|^2}$ and the phase delay between the oscillators $\psi = \arg [a_2^* a_1]$ \cite{SI}:
\begin{eqnarray}
\begin{aligned}
    \dot{A} &=& -\frac{\gamma}{2} A + g_a A \sin (\psi) f \left( A \sin \frac{\psi}{2}  \right), \\
    \dot{\psi} &=& \Delta \Omega - 4 g_a \sin^2 \left( \frac{\psi}{2} \right) f \left( A \sin \frac{\psi}{2}  \right).
\end{aligned}
\label{eq:analytic}
\end{eqnarray}
Here, $f(x) = 2 J_1(x)/x$ depends on the first-order Bessel function of the first kind $J_1$. The equations of motion \eqref{eq:analytic} are valid only in the $\mathcal{PT}$ symmetry-broken phase and have two steady-state solutions: (i) a collective state with vanishing oscillations ($A = 0$) but a stable phase delay $\psi=2\arcsin(\sqrt{\Delta\Omega/4g_a})$, and (ii) a coherently oscillating state with $A$ satisfying $f\left( A\Delta\Omega/\sqrt{\gamma^2+\Delta\Omega^2}\right)=(\gamma^2+\Delta\Omega^2)/4g_a\Delta\Omega$, and phase delay $\psi=2\arctan(\Delta\Omega/\gamma)$. As $f(x)\leq 1$, the two solutions exist in regions separated by the threshold defined by $g_a =  \left( \gamma^2 + \Delta \Omega ^2 \right)/ 4\Delta\Omega$. Above this threshold the first solution becomes unstable and the second, truly nonlinear solution -- a stable limit cycle -- emerges, thus revealing a Hopf bifurcation.
 
The dynamics of the collective motion under conditions defined by the two solutions of Eq.~\eqref{eq:analytic} are shown in the middle and bottom row of Fig.~\ref{fig2}, respectively, while the top row features the standard behavior of uncoupled thermal oscillators for comparison. In the linear regime, the individual particles' motion follows a Gaussian distribution, and we observe an increased motional amplitude as we transition into the $\mathcal{PT}$ symmetry-broken phase. For higher coupling rates, the particles' motion become nonlinear, which is reflected in a modified motional statistics to a displaced Gaussian distribution (Fig.~\ref{fig2}a). We compute the phases $\bar{\psi}_{1,2}$ of the individual particles' motion via the Hilbert transform and calculate the instantaneous phase delay between the oscillators as $\bar{\psi}= \bar{\psi}_{2} - \bar{\psi}_{1}$ \cite{SI}. The stable phase delay $\bar{\psi}$ is resolved in the zoomed-in time trace of the nonlinear motion (Fig.~\ref{fig2}b). The interaction generates a preferred phase delay $\bar{\psi}_\text{max}$ as observed in the histograms of $\bar{\psi}$, while it is uniformly distributed in the case of uncoupled oscillators.  The $z_1$-$z_2$ distribution shows the collective dynamics of the two particles (Fig.~\ref{fig2}c). In the case of uncoupled particles, the distribution is well-described by an uncorrelated two-dimensional Gaussian distribution. However, in the case of solution (i), $z_1$ and $z_2$ are strongly correlated, and thus the joint distribution is squashed under an angle that depends on $\bar{\psi}$. For nonlinear motion, the $z_1$-$z_2$ distribution exhibits a stable path -- the limit cycle. 

\begin{figure}[t!]
    \centering
            \includegraphics[width=\linewidth]{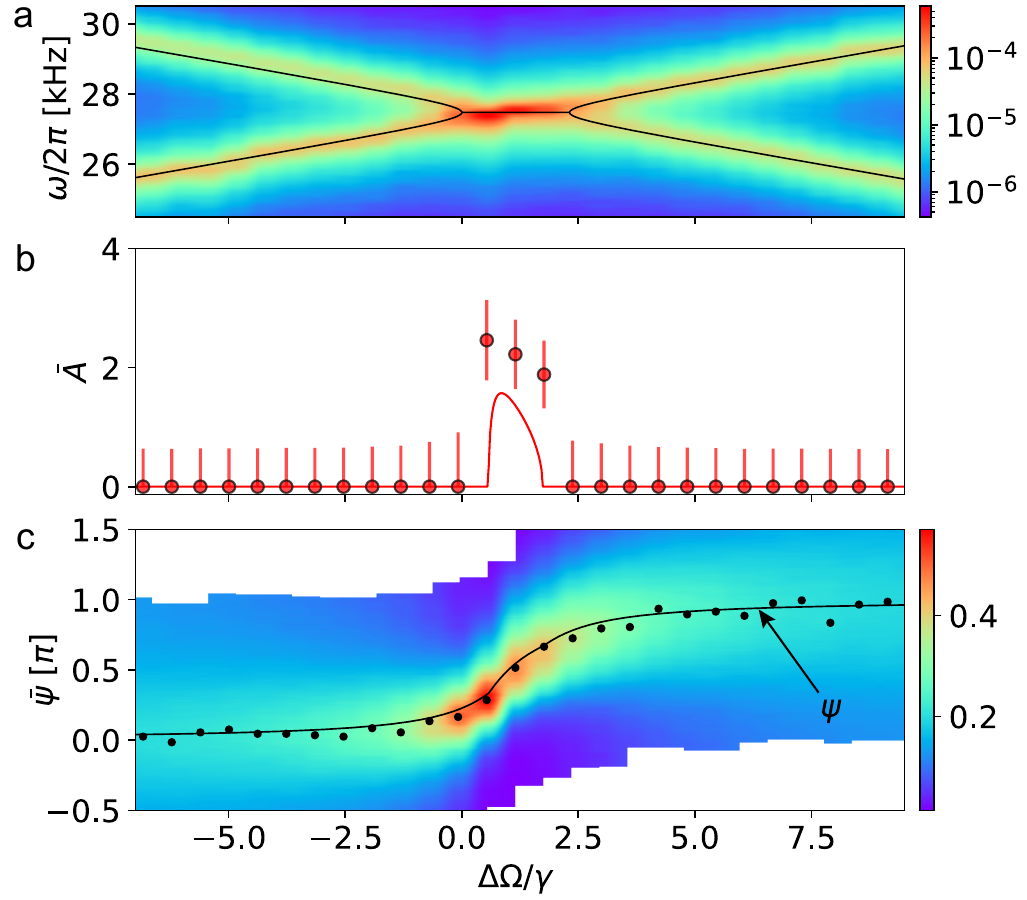}
    \caption{\textbf{Amplitude and phase delay of nonlinear motion.}
    \textbf{a.} The spectrogram of the particle motion reveals the mode eigenfrequencies as a function of $\Delta \Omega/\gamma$, where from the fit (black lines) we obtain a coupling of $g_a/\gamma = 0.58\pm 0.05$. The frequencies are degenerate for $0<\Delta\Omega/\gamma\lesssim 2.3$. \textbf{b.} Measured displacement amplitude $\bar{A}$ (points) and theoretical limit cycle amplitude $A$ (line)  as a function of $\Delta \Omega/\gamma$. The displacement amplitude becomes nonzero within the $\mathcal{PT}$ symmetry-broken phase whenever the particles' motion are limit cycles. \textbf{c.} Probability density of the phase delay $\bar{\psi}$ is shown in color, where black points show the obtained $\bar{\psi}_{\text{max}}$ at each detuning $\Delta\Omega/\gamma$. For large detunings $|\Delta\Omega/\gamma|\gg 0$, histograms of $\bar{\psi}$ follow an approximately uniform distribution. Within the $\mathcal{PT}$ symmetry-broken phase, $\bar{\psi}$ acquires a strongly preferred value that depends on $\Delta \Omega$. The solid black line is the theory plot that combines the linear and nonlinear models for $\psi $.}
    \label{fig3:detscan}
\end{figure}

We compare the theoretical steady-state solutions $A$ and $\psi$ to the measured displacement amplitude $\bar{A}$ and phase delay $\bar{\psi} $ as a function of the mechanical detuning $\Delta \Omega/\gamma$ (Fig.~\ref{fig3:detscan}). We obtain the coupling rate of $g_a/\gamma = 0.58\pm 0.05$ from the fit of the eigenfrequencies $\Omega_{\pm}$ to the spectrogram of particle motion (Fig.~\ref{fig3:detscan}a). The region between the EPs, given by $0\leq\Delta \Omega/\gamma\lesssim 2.3$, defines the $\mathcal{PT}$ symmetry-broken phase. To obtain the displacement amplitude $\bar{A}$ from the particle motion $z_{1,2}$, we first reconstruct the complex amplitudes $a_{1,2}$ \cite{SI}. We model the histograms of $|a_{1,2}|$ with Rice distributions and fit the individual displacements $\bar{A}_{1,2}$, which we use to calculate $\bar{A}=2k\sqrt{\hbar/m\Omega_0}  \sqrt{|\bar{A}_2|^2 + |\bar{A}_1|^2}$. Although there is a good quantitative agreement between the predicted $A$ (line) and reconstructed amplitudes $\bar{A}$ (points) (Fig.~\ref{fig3:detscan}b), we attribute the discrepancy between them to thermal fluctuations that are not included in our theory model. 
As the phase is periodic with $2 \pi$, we plot the histograms of $\bar{\psi} \mod 2\pi$ (colored density plot), centered around the preferred phase $\bar{\psi}_\text{max}$ (black points), as a function of $\Delta\Omega/\gamma$ (Fig.~\ref{fig3:detscan}c).  Throughout the scan, the phase difference undergoes a $\pi$ shift, as predicted by our model (black line) that combines the nonlinear theory in the range where $A \neq 0$ and the full linear theory with included thermal fluctuations elsewhere. 

\begin{figure}[t]
    \centering
    \includegraphics[width=\linewidth]{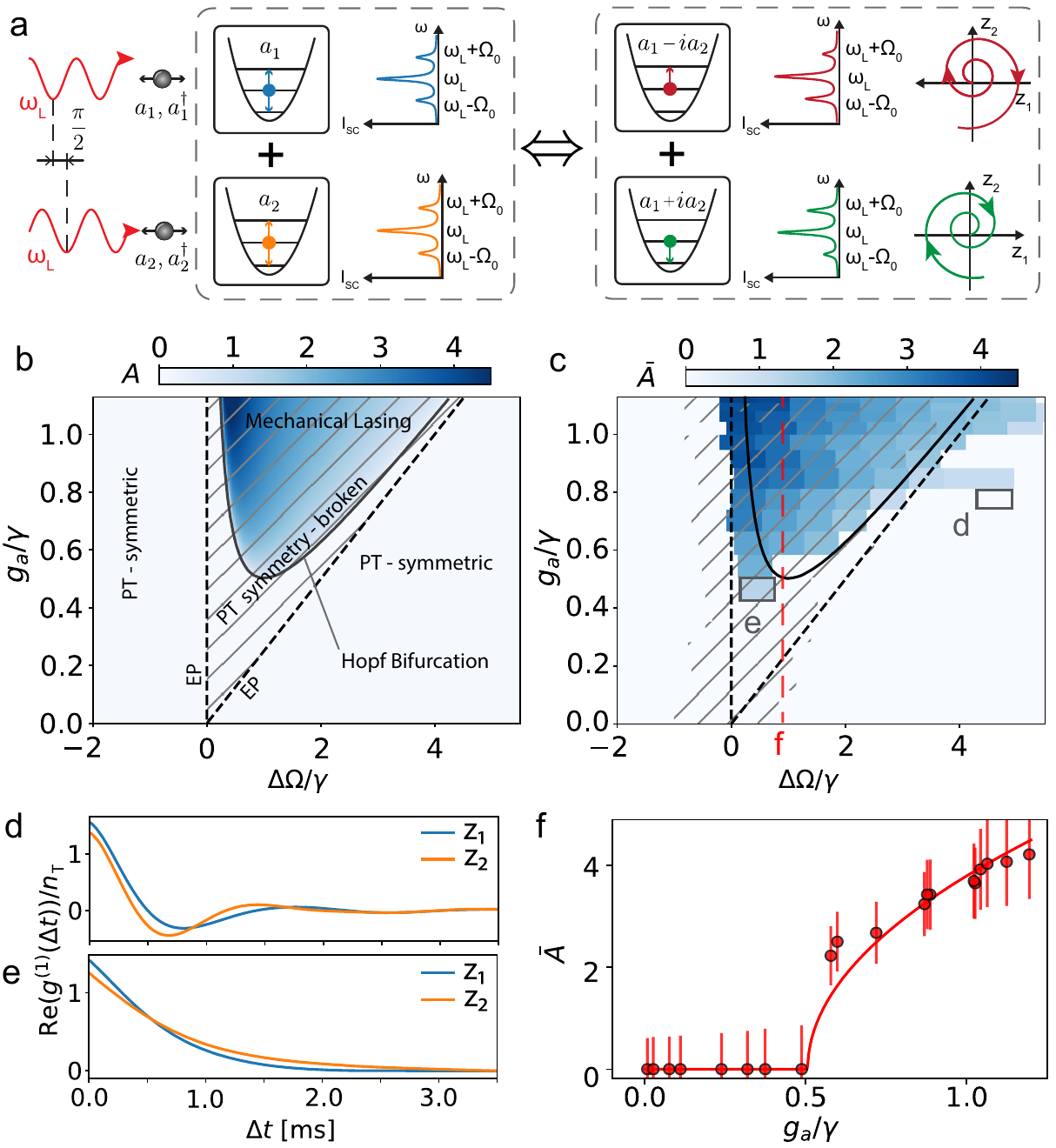}
    \caption{\textbf{Non-Hermitian phases.} \textbf{a.} The motion of particles 1 (blue) and 2 (orange) along the optical axes generates Stokes and anti-Stokes sidebands at $\omega_{\text{L}}-\Omega_0$ and $\omega_{\text{L}}+\Omega_0$, respectively, from the intrinsic laser frequency $\omega_{\text{L}}$. The optical interaction leads to modified amplitudes of Stokes and anti-Stokes sidebands of the eigenmodes $a_1-i a_2$ (red) and $a_1+i a_2$ (green), amplifying and damping the modes with suppressed anti-Stokes (red) and Stokes sideband (green), respectively. \textbf{b.} Expected amplitude $A$ of the limit cycle as a function of the detuning $\Delta \Omega/\gamma$ and the anti-reciprocal coupling $g_a/\gamma$. The oscillators are in the $\mathcal{PT}$ symmetry-broken phase (hatched region) between the exceptional points (dashed lines). Outside of this region, the oscillators are in the $\mathcal{PT}$-symmetric phase. The black line marks the Hopf bifurcation of the nonlinear model. The collective motion follows the limit cycles in regions with nonzero amplitude $A$ (shades of blue). \textbf{c.} Measured displacement amplitude $\bar{A}$, where each pixel represents a single measurement. \textbf{d.} (\textbf{e.}) The autocorrelation function $g^{(1)}(\Delta t)$ shows oscillatory (bi-exponential) behavior in the $\mathcal{PT}$-symmetric (symmetry-broken) phase. The shaded region in \textbf{b} marks the $\mathcal{PT}$ symmetry-broken phase if the measured autocorrelation function $g^{(1)}(\Delta t)$ shows a bi-exponential decay. \textbf{f.} The region of the limit cycle solutions ($ \bar{A}>0$) occurs at a threshold coupling rate of $g_a/\gamma>0.5$ for $\Delta \Omega/\gamma = 0.9$. The points are measured $\bar{A}$ and the line represents a theory fit. }
    \label{fig4:lasing}
\end{figure}

\paragraph{\textbf{Mechanical lasing transition.}} 
The Hopf bifurcation and the increase of the limit cycle amplitude $A$ may be interpreted as a mechanical lasing transition. In analogy to a laser, here the two-level laser medium is represented by the highly populated tweezer (upper level) and the Stokes sideband given by the particle motion (lower level). Scattering of the tweezer mode into the Stokes sideband creates a phonon in the mechanical degrees of freedom of the particles. In past works, one or more cavities were required to suppress the opposite transition into the anti-Stokes sideband \cite{Zhang2018}. Instead of enhancing the Stokes scattering via the Purcell effect, here an interference between the mechanical sidebands created by the particles' motion leads to a suppression of the Stokes or anti-Stokes scattering, depending on which mechanical mode is excited (Fig.~\ref{fig4:lasing}a). This interference leads to an amplified (suppressed anti-Stokes scattering) and a decaying mechanical mode (suppressed Stokes scattering). The lasing threshold is reached when the amplification exceeds the net loss, given by the intrinsic damping of the mechanical modes.

To fully characterize the mechanical lasing transition, we repeat the measurements of $\bar{A}$ as a function of $\Delta\Omega/\gamma$ for varying couplings $g_a/\gamma$. Comparison of $A$ from the nonlinear theory model in Eq.~\eqref{eq:analytic} (Fig.~\ref{fig4:lasing}b) to the extracted $\bar{A}$ (Fig.~\ref{fig4:lasing}c) shows a good agreement, which demonstrates that the observed nonlinear dynamics is well-described by our theory model. We distinguish three regions in both graphs. For $0 < \Delta \Omega/\gamma < 4 g_a/\gamma$ (hatched area) the oscillators are in the $\mathcal{PT}$ symmetry-broken phase, while they are in the $\mathcal{PT}$-symmetric phase elsewhere. Each particle motion is a combination of the two eigenmodes, which have non-degenerate frequencies outside and non-degenerate damping rates inside the $\mathcal{PT}$ symmetry-broken phase. Therefore, the crossing between these phases is characterized by the change of the amplitude correlation functions $g_{jj}^{(1)}(\Delta t ) =  \langle a_j^* (t) a_j (t+\Delta t) \rangle$ from an oscillatory behavior outside (Fig.~\ref{fig4:lasing}d), to a bi-exponential decay inside the symmetry-broken phase (Fig.~\ref{fig4:lasing}e). Furthermore, above the threshold marked by the solid black line, the oscillators exhibit the mechanical lasing transition of Eq.~\eqref{eq:analytic}. To clearly show the transition, we plot $\bar{A}$ as a function of the coupling at a detuning of $\Delta \Omega/\gamma \approx  0.9$ (Fig.~\ref{fig4:lasing}f). A nonzero $\bar{A}$ emerges at the threshold coupling of $g_{a}/\gamma \approx 0.5$ and follows a square-root dependence on the coupling, as predicted by the theory. The mechanical lasing transition could be understood as a second-order (non-Hermitian) phase transition \cite{Fruchart2021}, bearing in mind that our system does not exhibit an apparent thermodynamic limit. 

\paragraph{\textbf{Conclusion.}} We investigated the collective linear and nonlinear dynamics of two anti-reciprocally coupled optically levitated nanoparticles, which interact through light-induced dipole-dipole forces. The particles' motion was monitored independently, which allowed us to reconstruct the eigenmodes of the system and to observe signatures of the $\mathcal{PT}$ symmetry breaking. Within the $\mathcal{PT}$ symmetry-broken phase, two eigenmodes with different damping rates emerge and the particles' motion become strongly correlated, which we confirm by measuring a stable phase delay between the oscillators. For a sufficiently high coupling rate, the system shows a mechanical lasing transition where the joint phase space distribution exhibits a limit cycle. 

The presented steady-state measurements are a first step toward probing effects arising from the dynamical operation of nonreciprocally coupled particle chains induced by, for example, encircling the exceptional point for topological energy transfer \cite{XuHarris2016, Patil2022, Doppler}. Larger arrays of tunable, nonreciprocally interacting optically levitated particles have a great potential for studies of nonreciprocal phase transitions \cite{Fruchart2021}, nonequilibrium physics \cite{Loos2023, Xu2023, LiuPretherm}, and may have sensing applications \cite{McDonald2020, MalzMagnonAmplification, Porras2019, Mishkat, Wiersig2020, EPsReview}.
A combination of long-range nonreciprocal interactions and the already established quantum control of particle motion \cite{Delic2020} holds promise to be a game-changer for studying collective quantum phenomena, such as non-Hermitian quantum physics with long-range interactions \cite{LourencoLongRange}, entanglement in an optical cavity, and observation of quantum optical binding \cite{Rudolph2023}.

\begin{acknowledgments}
	\textbf{\textit{Acknowledgments.}} We thank Tobias Donner for insightful discussions, Markus Aspelmeyer for his support, Helmut Ritsch for his idea of how to calibrate the particle distance, and Baptiste Courme and Iurie Coroli for their help with programming the acousto-optical deflectors. This research was funded in whole or in part by the Austrian Science Fund (FWF, Project No. I 5111-N). A.V.Z. acknowledges support from the European Union’s Horizon 2020 research and innovation programme under the Marie Sklodowska-Curie grant LOREN (grant agreement ID: 101030987). M.A. acknowledges support from the European Union's Horizon 2022 research and innovation programme under the Marie Sklodowska-Curie grant NEOVITA (grant agreement ID: 101109773). H.R., K.H., B.A.S. acknowledge funding from the Deutsche Forschungsgemeinschaft (DFG, German Research Foundation) under Grant No. 439339706. B.A.S. acknowledges support by the Deutsche Forschungsgemeinschaft (DFG, German Research Foundation) under Grant No. 510794108. For the purpose of Open Access, the author has applied a CC BY public copyright license to any Author Accepted Manuscript (AAM) version arising from this submission. 
\end{acknowledgments}

\textbf{\textit{Note.}} We are aware of related work by Li\v{s}ka \textit{et al} \cite{Liska2023}.

\bibliographystyle{apsrev4-1}

\appendix
\makeatother

\setcounter{figure}{0}
\setcounter{equation}{0}
\renewcommand{\thefigure}{S\arabic{figure}}
\renewcommand{\theequation}{S\arabic{equation}} 

\clearpage
\onecolumngrid
\appendix
\pagenumbering{alph}

\section{\large Supplementary Materials}

\subsection{Theoretical model}
In this section, we discuss the derivation of the linear and nonlinear models \eqref{eq:equmotion} and \eqref{eq:analytic} for the collective motion of the two trapped nanoparticles from the theory of optical interactions \cite{Dholakia2010,Rudolph2023}. The starting point is the quantum Langevin equations from Ref.~\cite{Rudolph2023} for light-induced dipole-dipole interactions between arbitrary mechanical degrees of freedom. We consider two identical spherical particles in their respective tweezer traps (see Eq.~(69) \cite{Rudolph2023}), ignore quantum noise, treat all operators as classical variables, and add gas damping and diffusion with damping constant $\gamma$ and gas temperature $T_{\rm g}$. The tweezers are assumed to have identical (linear) polarizations and waists but may be driven with a different power. The mechanical frequencies along transverse directions are significantly higher than the frequency along the optical axis, therefore we restrict the dynamics to the particle motion along the tweezer propagation direction.


To determine the main nonlinearity of the two-particle dynamics around the tweezer foci, we expand the optical potential and the non-conservative radiation pressure force to the first-order nonlinearity, while in the dipole-dipole interaction, the tweezer fields are well-approximated by plane waves. Furthermore, the equations of motion are expanded to leading order in the inverse distance $1/d_0$ between the tweezer foci, since far-field coupling dominates.

The resulting equations of motion for the particle coordinates $z_{1,2}$ and momenta $p_{1,2}$ along the optical axis can be written in terms of the complex amplitudes $a_{1,2}$, as defined below of Eq.~\eqref{eq:equmotion}. For the experiment described in this article, the influences of gas collisions, nonlinearities in the optical potential, the non-conservative radiation pressure forces, dipole-dipole interactions as well as the mechanical frequency difference $\Delta\Omega$ are weak during one cycle of the mean harmonic motion $\Omega_0$. This allows us to average the equations of motion for $a_{1,2}$ over one mechanical period $2\pi/\Omega_0$, resulting in
\begin{equation}\label{eq:nonlinearcomplex}
    \dot a_j = \pm i \frac{\Delta\Omega}{2}a_j - \frac{\gamma}{2}a_j + i\beta |a_j|^2 a_j + \sqrt{\gamma n_T} \xi_j(t)
    \pm i (g_r \pm g_a) (a_2 - a_1) f(2 k z_{\rm zpf, 0} |a_2 - a_1|),
\end{equation}
where the upper (lower) sign holds for $j=1$ $(j=2)$. Here, we define $z_{\rm zpf, 0} = \sqrt{\hbar/2m\Omega_0}$, the thermal occupation $n_T = k_{\rm B}T_{\rm g}/\hbar\Omega_0$, the optical potential nonlinearity $\beta = 3\Omega_0 z_{\rm zpf, 0}^2/z_{\rm R}^2$, depending on the Rayleigh range $z_{\rm R}$ of the tweezers, and the reciprocal and anti-reciprocal coupling rates $g_r = G \cos{(k d_0)} \cos{(\Delta \phi)} / k d_0$ and $g_a = G \sin{(k d_0)} \sin{(\Delta \phi)} / k d_0$ as in the main text. They depend on the constant
\begin{align}
    G = \frac{\varepsilon_0\chi^2 V^2 k^5 E_0^2}{16\pi m \Omega_0}\cos^2\theta,
\end{align}
with the particle volume $V$ and electric susceptibilty $\chi = 3(\varepsilon_r - 1)/(\varepsilon_r + 2)$ with relative electric permeability $\varepsilon_r$, the local tweezer field strength $E_0$ and the angle ($\pi/2-\theta$) between tweezer polarization and particle-connecting axis. Importantly, the relevant nonlinearity in Eqs.~\eqref{eq:analytic},~\eqref{eq:nonlinearcomplex} causing the limit cycle is \textit{not} the optical potential nonlinearity $\beta$, but the nonlinearity in the dipole-dipole interaction described by the function $f$. As a result of the average over one mechanical cycle, (i) the non-conservative radiation pressure forces cancel, and (ii) the gas diffusion noises $\xi_j(t)$ turn complex-valued.

{\it First order correlation functions.--} To obtain the linear model \eqref{eq:equmotion} for deeply trapped particles, we harmonically expand Eq.~\eqref{eq:nonlinearcomplex} around $a_{1,2}=0$, using $f(0) = 1$. Then, the autocorrelation functions $g^{(1)}_{jj}=\langle a^*_j(t)a_j(t+\Delta t)\rangle$ for $j\in\{1,2\}$, as well as the cross-correlation function $g^{(1)}_{12}=\langle a^*_2(t)a_1(t+\Delta t)\rangle$ of deeply trapped particles follow from a straightforward calculation, yielding
\begin{align}\label{eq:autocorrelation}
    & g_{jj}^{(1)}(\Delta t) = n_T \Bigg[ \Bigg( 1 + \frac{4g_a(g_a\pm g_r)}{\gamma^2 + \kappa^2} \Bigg)\cos\left( \frac{\kappa}{2}\Delta t\right)+ \frac{4\gamma g_a(g_a\pm g_r)}{\kappa(\gamma^2 + \kappa^2)} \sin\left(\frac{\kappa}{2}|\Delta t| \right)  \pm i \frac{\Delta\Omega - 2g_a}{\kappa}\sin\left(\frac{\kappa}{2}\Delta t\right)\Bigg]e^{-ig_r \Delta t } e^{-\frac{\gamma}{2}|\Delta t|},\\
    \label{eq:crosscorrelation}
    & g_{12}^{(1)}(\Delta t) = n_T \Bigg[ \frac{2g_a(i\gamma+ 2 g_a-\Delta\Omega)}{\gamma^2 + \kappa^2}  \Bigg\{
    \cos\left( \frac{\kappa}{2}\Delta t\right)  + \frac{\gamma}{\kappa} \sin\left(\frac{\kappa}{2}|\Delta t| \right) \Bigg\} 
    +i \frac{2g_r}{\kappa}\sin\left(\frac{\kappa}{2}\Delta t\right)\Bigg]e^{-ig_r \Delta t } e^{-\frac{\gamma}{2}|\Delta t|},
\end{align}
with $\kappa = \sqrt{\Delta\Omega^2 - 4g_a\Delta\Omega + g_r^2}$. In the $\mathcal{PT}$ symmetry-broken regime, where $\kappa$ becomes imaginary, the oscillatory behavior in Eq.~\eqref{eq:autocorrelation} changes to a bi-exponential decay, in excellent agreement with the observed correlation functions (Fig.~\ref{fig4:lasing}d-e).

{\it Limit cycle model.--} The limit cycle model \eqref{eq:analytic} follows from Eq.~\eqref{eq:nonlinearcomplex} by ignoring gas diffusion, choosing the anti-reciprocal interaction as $g_r=0$ and writing the complex amplitudes in terms of occupations and phases as $a_{1,2} = \sqrt{n_{1,2}}\exp{(-i\psi_{1,2})}$. It follows that the occupation difference $n_2 - n_1$ is always damped to zero on the timescale of the gas damping constant $\gamma$, while both occupations $n_{1,2}$ and the mechanical phase delay $\psi = \psi_2 - \psi_1$ decouple from the mean mechanical phase $(\psi_1 + \psi_2)/2$. Assuming that $n_2-n_1$ has already settled at zero, and defining the dimensionless effective oscillation amplitude by $A = 2\sqrt{2}kz_{\rm zpf, 0}\sqrt{n_1 + n_2}$, we finally arrive at Eq.~\eqref{eq:analytic}. Note that the optical potential nonlinearity $\beta$ cancels from the effective limit cycle model \eqref{eq:analytic}, as its influence on the phase delay $\psi$ vanishes if the particles oscillate with identical amplitude.

\subsection{Detection of the particle motion}

\begin{figure}[ht]
    \centering
    \includegraphics[width=0.5\linewidth]{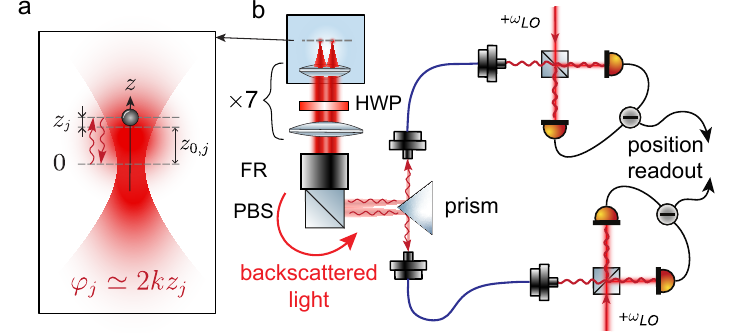}
    \caption{\textbf{Detection of the particles' motion.} \textbf{a.} A particle is trapped in a tightly focused Gaussian beam at a position $z_{0,j}$ away from the focus. The phase of the backscattered light changes with the particle motion $z_j$ around the trap position as $\varphi_j \approx 2kz_j$. \textbf{b.} Sketch of the balanced heterodyne detection scheme. The backscattered light is separated from the incoming beams by a Faraday rotator (FR) and a polarizing beamsplitter (PBS) and subsequently reimaged onto a prism mirror to spatially separate the light scattered by the two particles. Each particle's image is collected with a single-mode fiber in a confocal microscope configuration for further spatial filtering from the background reflections. The particles' signals are combined with a local oscillator at a frequency of $\omega_{\text{LO}}$ with a 50:50 beamsplitter and are measured with balanced photodetectors.}
    \label{figM:Det}
\end{figure}

The particles scatter light and imprint the information about their position onto the phase of the scattered light \cite{Tebbenjohanns2019}. In our case, the optical phase of the backscattered light optimally maps the particles' motion along the optical axis (Fig.~\ref{figM:Det}a). The optical phase $\varphi_j$ can be approximated as the sum of contributions from the three-dimensional motion:
\begin{equation}
    \varphi_j(r_j(t)) = \varphi_j( z_j) +\varphi_j(y_j) + \varphi_j(x_j).
\end{equation}
However, the $z$-motion is well-separated in frequency from the $x$- and $y$-motion due to the trap asymmetry ($\Omega_{z,j}\ll \Omega_{x,j},\Omega_{y,j}$). Therefore, we can isolate the $z$-motion contribution by applying a bandpass filter with the bandwidth $[5,50]$ kHz around the average $z$ frequency $\Omega_0/2\pi=27.5$ kHz. Furthermore, the transverse motion influences the optical phase around $100$ times less than $z_j$ in the backplane detection. 

To split the backscattered light from the trapping light, we use a polarizing beamsplitter and a Faraday rotator (Fig.~\ref{figM:Det}b). The image of the particle plane is magnified $160$-fold onto a prism mirror that separates the individual particle images into opposite directions. The split particle signals are collected with two single-mode fibers such that the confocal microscope magnification ($\times 7$) between the trapping plane and the fiber input plane maximizes the collection efficiency \cite{Magrini2021, Tebbenjohanns2021}. The collected scattered light (optical power of $5$-$10$ $\mu$W) is mixed on a beamsplitter with a reference beam (local oscillator) (optical power of $1$-$2$ mW, frequency-shifted by $\omega_{\text{LO}}/2\pi=1.1$ MHz) to implement a balanced heterodyne detection.


The heterodyne signal to the leading order in the particle motion $z_j$ yields:
\begin{equation}
    I_{-,j}(t)\propto \left|E_{LO}^*E_{sc}\right|\cos\left(\omega_{LO}t+\varphi_j[z_j(t)]\right),
\end{equation}
where $E_{LO}$ and $E_{sc}$ are the strengths of the local oscillator and scattered fields, respectively.
We retrieve the optical phases $\varphi_j$ by calculating the Hilbert transform of $I_{-,j}$, which allows us to obtain the unwrapped argument of $I_{-,j}$. Finally, we subtract the known accumulated phase of the local oscillator to obtain:
\begin{equation}
    \varphi_j(z_j(t)) = \textrm{unwr}\left\{ \textrm{arg}\left( \mathcal{H}\left[I_{-,j}(t)\right] \right)\right\} -\omega_{LO}t.
\end{equation}
The obtained phase can now be converted into the position $z_j$ as \cite{Tebbenjohanns2019}:
\begin{equation}
    \label{eqM:gouy}
    \varphi_j(z_j)=2kz_j-\textrm{arctan}\left((z_j + z_{0,j})/z_{\rm R}\right),
\end{equation}
where $\textrm{arctan}\left((z_j + z_{0,j})/z_{\rm R}\right)$ is the modification due to the Gouy phase with the constant offset from the focus $z_{0,j}$, the Rayleigh length $z_{\rm R}=w_{x}w_{y}\pi/\lambda$ and waists $w_{x}$ and $w_{y}$ along the $x$- and $y$-axis, respectively. We numerically calculate the tweezer waists $w_{x} = 0.678 \ \mu $m and $w_{y} = 0.775 \ \mu $m from the measured mechanical frequencies and $z_{0,j}=0.8 \ \mu$m from the particle radius to estimate the correction due to the Gouy phase. The obtained Rayleigh length of $z_{\rm R}  = 1.55 \ \mu $m results in the modification of $\sim 3 \% $ to the particle motion detection. Therefore, we safely omit this correction and use $\varphi_j ( z_j) = 2 k z_j$ instead in the main text.


\subsection{Comparison of particle sizes}

\begin{figure}[h]
    \centering
    \includegraphics[width=0.4\linewidth]{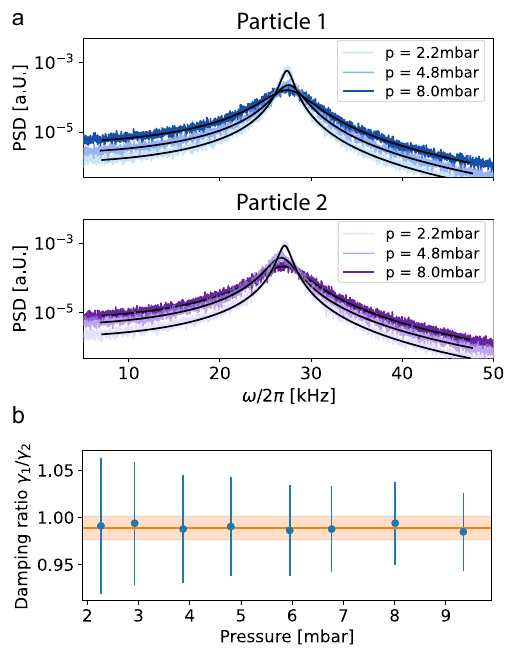}
    \caption{\textbf{Ratios of particle damping rates at different pressures.} \textbf{a.} Calculated particle PSDs at three different pressures for particles 1 (blue) and 2 (purple) illustrate how the damping rates $\gamma_{1,2}$ decrease as the gas pressure decreases. The black lines correspond to the fitted lineshapes.
    \textbf{b.} Ratios of the fitted damping rates at different pressures (blue points and error bars) yields the weighted average (orange line) and its standard deviation (orange shaded region) of $\gamma_{1}/\gamma_{2}=0.99 \pm 0.01$.}
    \label{fig:sup:particlesize}
\end{figure}

We assume identical spherical particles in size and mass in the theoretical models. 
The particles used in the experiment have a nominal radius of $r = (105 \pm 2)$ nm (microParticles GmbH), therefore they could be of slightly different sizes. 
In the regime dominated by gas damping, the damping rate of the particle CoM motion depends on the particle radius as $\gamma\propto 1/r$ \cite{Gieseler2013}. Therefore, we measure and compare the damping rates of the two particles as a function of gas pressure to determine the ratio of the particle radii \cite{Yan2023}. At each pressure, we calculate the PSDs of the CoM motion and extract the gas damping rates of particles 1 (blue) and 2 (purple) from the fit of the driven damped harmonic oscillator (Fig.~\ref{fig:sup:particlesize}a) \cite{Cavoptomechanics}. We assume equal mass densities and, due to the absence of rotational signatures in the PSDs, spherical particles. 
This relates the ratio of the damping rates of the two particles simply to the inverse ratio of the particle radii $\gamma_{1}/\gamma_{2}=r_2/r_1$.  
We obtain the average ratio of the damping rates $\gamma_{1}/\gamma_{2}=0.99 \pm 0.01$, and determine the size ratio to be $r_1 / r_2 = 1.01 \pm 0.01$ (Fig.~\ref{fig:sup:particlesize}b). We conclude that the particles can be considered of approximately equal sizes.

\subsection{Calibration of the particle distance}

\begin{figure}[h]
    \centering
    \includegraphics[width = 0.5\linewidth]{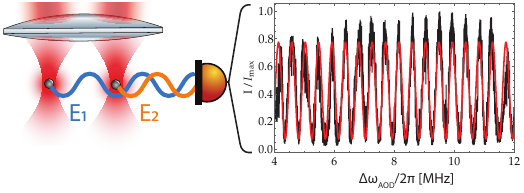}
    \caption{\textbf{Distance calibration.} The scattered light from particles 1 and 2 interferes in the far field. We measure the intensity of the interference pattern while scanning the particle distance (black line). The fringe spacing is given by the wavelength of the trapping light. We fit the pattern (red line) to determine the calibration constant from the frequency difference to particle distance.}
    \label{figM:distance}
\end{figure}

The mean particle distance $d_0$, given by the positions of the foci of their respective tweezers, influences the reciprocal and anti-reciprocal coupling rates. In our experiment, we tune the frequency difference $\Delta \omega$ of the RF tones to modify the distance as $d_0=C_0 \Delta\omega$, where $C_0$ is the conversion factor that can be calibrated by comparing the distance to a known etalon, e.g. the laser wavelength. Therefore, we perform an interferometric measurement of the dipole radiation from the two particles with a detector placed in the far field, along the axis connecting the particles (Fig.~\ref{figM:distance}). The tweezer polarizations are set perpendicular to this axis, such that the dipole radiation in the direction of the detector is maximal. As the images of the two particles are indistinguishable at the detector, the overlap of the dipole radiations leads to an interference that depends on $d_0$. In our measurement, we sweep $\Delta\omega/2\pi$ in the range of $4-12$ MHz and record the intensity of the interference pattern. We note that the magnitude of the interference pattern changes as a function of $\Delta\omega$. As the measurement is fully repeatable, we attribute this to the varying AOD diffraction efficiencies as a function of the RF tone frequencies. We fit the data with a sine function and obtain the conversion factor $C_0 = (2.3845 \pm 0.0005) \: \mu$m/MHz. All measurements in the main text were performed at a frequency difference of $\Delta \omega / 2 \pi = 7.72$ MHz, which leads to an estimated distance of $d_0 = (18.408 \pm 0.005) \: \mu$m.

\subsection{Measurement of the coupling rates}

In this section, we detail the experimental methods to optimize the anti-reciprocal coupling rate $g_a$ and to reduce the reciprocal coupling rate $g_r$ and the coupling rate $g_C$ due to electrostatic interactions.

{\it Optical coupling---} The optical coupling rates $g_r$ and $g_a$ depend on the particle distance $d_0$ and the optical phase difference $\Delta \phi$. 
To maximize $g_a$ and minimize $g_r$, we scan both parameters while the trap polarization is set to $\sim 0^\circ$ to maximize the optical interaction. For a general combination of parameters where $g_a \neq 0 $ and $g_r \neq 0$, the total coupling rates $g_r\pm g_a$ are different in opposite directions. In this case, the response to the motion of particle 2 in the PSD of particle 1, with a peak amplitude $\propto (g_r+g_a)^2$ at frequency $\Omega_2$ in the PSD, is different from the peak amplitude $\propto (g_r-g_a)^2$ of the particle 1 at frequency $\Omega_1$ in the PSD of particle 2. For any combination of $d_0$ and $\Delta \phi$, we monitor the PSD of both particles and we choose the parameters such that the two contributions have matching amplitudes, which is satisfied either for $g_r = 0$ or $g_a = 0$. As the final step, we make sure that there is no normal mode splitting but normal mode attraction to confirm that $g_r\equiv 0$. For such a configuration, we estimate that $\Delta \phi = (0.5 \pm 0.08)\:\pi$ due to the step size of the scan. With the previously estimated distance of $d_0 = (18.408 \pm 0.005) \: \mu$m, the reciprocal coupling $g_r\propto \cos(k d_0)\cos(\Delta \phi) = 0.00 \pm 0.08$ is effectively zero. 

\begin{figure}
    \centering
    \includegraphics[width=0.5\linewidth]{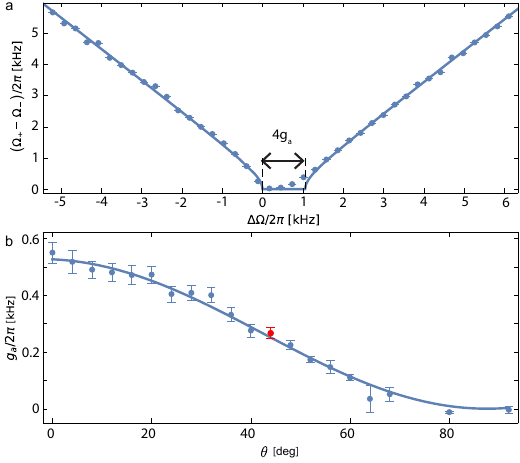}
    \caption{\textbf{Anti-reciprocal coupling rate. a.} We fit the eigenfrequency difference $\Omega_+ - \Omega_-$ as a function of the intrinsic frequency detuning $\Delta \Omega$ (points) with linear theory (line) to extract the coupling rate $g_a$. \textbf{b.} The anti-reciprocal coupling rate follows a $\cos^2 \theta$ dependence of the tweezer polarization angle $\theta$. The red point represents the detuning scan in \textbf{a}.}
    \label{figM:couplingrates}
\end{figure}

To determine the magnitude of the anti-reciprocal coupling rate $g_a$, we fit the difference of the eigenfrequencies $\Omega_+ - \Omega_-=\Re ( \sqrt{\Delta \Omega ^2 - 4 g_a \Delta \Omega } )$, obtained from the PSDs, as a function of the intrinsic frequency detuning $\Delta\Omega$ (Fig.~\ref{figM:couplingrates}a). We attribute the nonzero values in the $\mathcal{PT}$ symmetry-broken phase to small differences in the particle masses and the thermal fluctuations, which are not accounted for in the linear model of the eigenfrequencies. We determine $g_a$ as a function of the tweezer polarization $\theta$ (Fig.~\ref{figM:couplingrates}b), which follows a $\cos ^2\theta$ function, as predicted from the light-induced dipole-dipole interactions.

\begin{figure}
    \centering
    \includegraphics{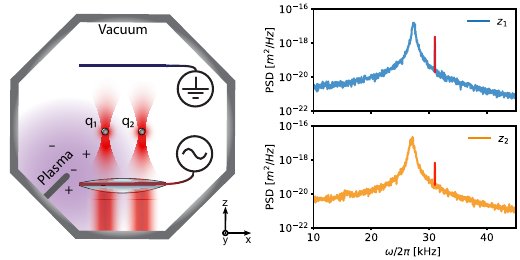}
    \caption{\textbf{Charge control and calibration.} \textbf{a.} To change the amount of net charges attached to the particles we ignite a plasma at a high-voltage pin inside the vacuum chamber. We monitor the number of charges by driving the particles via an AC field, applied to electrodes mounted to the trapping lens (signal, red line) and a lens tube (ground, dark blue line), at a relative distance of $D_e = (5 \pm 1)$ mm. \textbf{b.} The electric drive can be seen as a sharp peak (red) in the power spectral densities if the particles are charged. We estimate the remaining charges to be $q_1 = (2.5 \pm 0.7)\:e$ and $q_2 = (1.5 \pm 0.5)\:e$, where $e$ is the elementary charge.}
    \label{figM:charges}
\end{figure}

{\it Electrostatic coupling---} To minimize the electrostatic interaction, we discharge the particles by igniting a plasma at a high-voltage pin inside the vacuum chamber (Fig.~\ref{figM:charges}a). We attach wires to the holders of the trapping lens and the lens that collects light after the tweezer foci, which forms a capacitor for the charged particle with holders acting as the electrodes. We continuously drive the particle motion with a sinusoidal signal applied to two electrodes, which yields a sharp peak at the drive frequency in the PSD of the particle motion. When plasma is ignited, the number of charges on each particle fluctuates, which we monitor as changes in the peak amplitude in real-time. We switch off the high voltage when both particles have just a few charges, which is reflected in low amplitudes of the drive peak in the PSD.

To estimate the electrostatic interaction after the discharging procedure, we follow the procedure in the supplementary information of \cite{Rieser2022}. To calibrate the number of charges on each particle, we drive the motion of both particles by applying a sinusoidal voltage with an amplitude of $V_{AC} = 10$ V and at a frequency $\Omega_d/2\pi = 31$ kHz to the electrodes and record the particle motion. Assuming a model of a point charge in a plate capacitor, we estimate the number of charges $N_{1,2}$ to be: 
\begin{equation}
    \label{eqM:numbercharges}
    N_{1,2} = \frac{\sqrt{2} m D_e}{e V_{AC}} \left|\Omega ^2 _{1,2} - \Omega ^2 _d\right| \sqrt{\left\langle z^2_{d1,2} \right\rangle},
\end{equation}
where $m$ is the particle mass, $D_e = (5 \pm 1)$ mm is the distance between the electrodes, $e$ is the elementary charge, $\Omega_{1,2}$ are the oscillation frequencies of particles 1 and 2, and $\langle z^2_{d1,2} \rangle$ is the time-averaged displacement of particle 1,2 along the $z$-axis due to the electronic drive. The PSDs of the driven motion can be seen in Fig.~\ref{figM:charges}b. The peak due to the electric drive is marked in red. We estimate the number of charges to be $N_1 = 2.5 \pm 0.7$ and $N_2 = 1.5 \pm 0.5$, which gives an electrostatic coupling rate of:
\begin{equation}
    \label{eqM:coulomb}
    g_C = -\frac{N_1 N_2 e^2 }{8 \pi \varepsilon _0 m \Omega ' d^3_0} = (-0.077 \pm 0.030)\:\text{Hz}.
\end{equation}
Here, $\varepsilon _0$ is the vacuum permittivity and $\Omega ' = \sqrt{\Omega ^2 - N_1 N_2 e^2/(4 \pi \varepsilon_0 m d^3_0)}$. This coupling rate is more than two orders of magnitude smaller than the intrinsic damping rate and the measured anti-reciprocal coupling rates for polarization angles in the range $[0^\circ,80^\circ]$. Therefore, the electrostatic interaction does not affect the non-Hermitian dynamics in our experiment.

\subsection{Evaluation of the phase delay and the displacement amplitude}

{\it Phase delay.--} To compute the oscillation phases $\bar{\psi}_{1,2}$, we first apply the Hilbert transform to the individual particle motion $z_{1,2}$ to obtain the complex amplitude $\tilde z_{1,2}(t)=\mathcal{H}\left[z_{1,2}\right]$ such that $z_{1,2}(t)=\Re (\tilde z_{1,2}(t))$. The oscillation phases are recovered from the unwrapped phases of the complex amplitudes:
\begin{equation}\label{eq:mechphase}
    \bar{\psi}_j(z_j(t)) = \textrm{unwr}\left\{ \textrm{arg}\left( \mathcal{H}\left[z_{j}(t)\right] \right)\right\}.
\end{equation}
We define the instantaneous phase delay between the oscillators as $\bar{\psi} = \bar{\psi}_2 - \bar{\psi}_1$. The interaction generates a preferred phase delay $\bar{\psi}_{max}$ as observed in the histograms of $\bar{\psi}$, while it is uniformly distributed in the case of uncoupled oscillators (Fig.~\ref{fig2}b).

\begin{figure}
    \centering
    \includegraphics[width=0.5\linewidth]{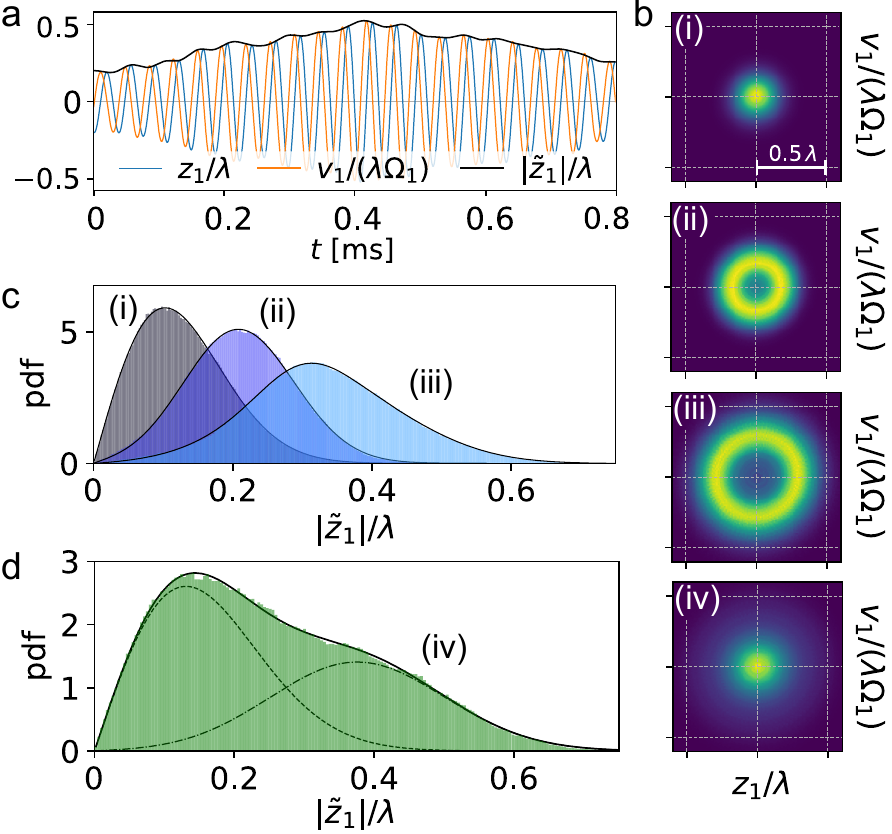}
    \caption{\textbf{Evaluation of the displacement amplitude.} \textbf{a.} Short time trace of the particle motion (blue), velocity (orange), and the reconstructed oscillation envelope (black).
    \textbf{b.} Phase space distributions of the motion of particle 1 sampled over the whole time trace for four different configurations, labeled as (i)-(iv). The limit cycle is observed in (ii) and (iii), while no limit cycle is observed in (iv).
    \textbf{c.} The probability distribution functions (pdf) of the oscillation envelopes for configurations (i)-(iii). Solid black lines represent the fits of the Rice distribution, where (ii) and (iii) clearly show the increase of the displaced amplitude $\bar A_1$ from the case of $\bar A_1=0$ in (i). The presented experimental data corresponds to the following detuning and coupling rate $(\Delta\Omega/2\pi,g_a/2\pi)$ and extracted displacement amplitudes $\bar A_1$: (i) $(4.68, 0.01)$ kHz and $0$, (ii) $(0.45, 0.29)$ kHz and $1.7$, (iii) $(0.79, 0.51)$ kHz and $2.7$. \textbf{d.} Histogram of $\tilde z_1/\lambda$ in the vicinity of the NIR region for $(\Delta\Omega/2\pi,g_a/2\pi)=(0.56, 0.51)$ kHz, corresponding to the distribution (iv), exhibits two peaks. The solid black line is the bi-modal Rice distribution fit, while the dashed and dash-dotted lines show the contributions of two Rice distributions with fitted $\bar A_1 = 0$ and $\bar A_1 = 3.1$, respectively.}
    \label{figM:Rice}
\end{figure}

{\it Displacement amplitude.--} The Nonlinear Interaction Regime (NIR) arises for the negative effective damping $\gamma_-<0$ as predicted by the linearized theory. This results in the modified statistics of the particles' motion and the emergence of a limit cycle (Fig.~\ref{fig2}a). To analyze this regime quantitatively and to compare the results to the theoretical model \eqref{eq:analytic}, we calculate the histograms of the oscillation envelopes $|\tilde z_j|$ for each particle and fit them with a model accounting for the possible emergence of a limit cycle with a displacement amplitude $\bar A_j$ (Fig.~\ref{figM:Rice}a) \cite{kepesidis2016}. 

Outside the NIR, there is no limit cycle, therefore $\bar A_j=0$. There, the positions (as well as velocities) assume Gaussian distributions centered around zero with a width $\sigma$ that depends on the effective thermal occupation.
The histograms of $|\tilde z_j(t)|$ are described by the Rayleigh distribution $P_{\rm Ray}(x, \sigma) = (x/\sigma^2) \exp( -x^2/ 2\sigma ^2  )$ as a function of the position $x$.
As soon as the limit cycle trajectories emerge, the displaced amplitude $\bar A_j$ becomes nonzero. 
There, the oscillation envelope follows a Rice distribution that generalizes the Rayleigh distribution for the case of non-centered Gaussian random variables. The Rice distribution has previously been used for the statistical description of precondensed light \cite{Santic_2018}. In our case, it is defined as:
\begin{equation}
    P_{\rm Rice}(x, \sigma, \bar A_j) = \frac{x}{\sigma ^2} e^{ - \frac{ x^2 + \bar A_j ^2 } { 2 \sigma ^2 } } \mathcal{I}_0 \left( \frac{ x \bar A_j }{ \sigma^2 } \right),
    \label{eq:me:rice}
\end{equation}
where $\mathcal{I}_0$ is the modified Bessel function of the first kind with order zero. Note that the thermal (coherent) statistics are recovered from the Rice distribution in the limit of $\bar{A}_j \rightarrow 0$ ($\bar{A}_j \gg \sqrt{\sigma}$).

We fit the Rice distribution to the histograms of the envelope $|\tilde z_j|$ with $\bar A_j$ and $\sigma$ as free parameters. 
The fitted amplitudes are then used to calculate the collective displaced amplitude $\bar A=2k\sqrt{\hbar/(m\Omega_0)}\sqrt{\bar A_1^2+\bar A_2^2}$ and compared to the theoretical amplitude $A$ from Eq.~\eqref{eq:analytic}.
Fig.~\ref{figM:Rice}c shows the histogram of the oscillation envelope for particle 1 fitted with a Rice distribution outside the NIR ((i)) as well as within the NIR for increasing interaction strength ((ii) and (iii)). In Fig.~\ref{figM:Rice}b we plot the corresponding position-velocity distributions of particle 1. In the vicinity of the NIR region, we observe a bi-modal distribution (Fig.~\ref{figM:Rice}d), which can be interpreted as a time average of the collective state occasionally jumping between the linear and nonlinear states. We attribute that to the thermal fluctuations-induced drifts of the mechanical frequencies around their average values. As a result, if the average frequencies are at the boundary of the limit cycle region, the particles spend some time in the limit cycle phase ($\bar A_j>0$) and the rest in the correlated thermal state  ($\bar A_j=0$) if otherwise. In these cases, we fit the histograms of the oscillation envelopes with a weighted sum of two different Rice distributions:
\begin{equation}
    P_{j}(x)=\alpha P_{\rm Rice}(x,\sigma_{j,1},\bar A_{j,1})
    +(1-\alpha) P_{\rm Rice}(x,\sigma_{j,2},\bar A_{j,2}).
\end{equation}
For example, at a detuning of $\Delta\Omega/2\pi=(0.56\pm 0.15)$ kHz and a coupling rate of $g_a/2\pi=(0.51\pm 0.03)$ kHz we obtain the amplitudes $\bar A_{1,1}=0$ with the weight $\alpha=0.56$ and $\bar A_{1,2}=3.1$ with the weight $1-\alpha=0.44$, thus confirming that our model explains the average collective dynamics well. In the case of a bimodal distribution for the individual oscillation envelope, we use the fitted displaced amplitude with the highest weight for the calculation of the collective displaced amplitude, shown in Fig.~\ref{fig4:lasing} of the main text.

\subsection{Experimental evaluation of the correlations}

\begin{figure*}[ht!]
   \centering
    \includegraphics[width=\linewidth]{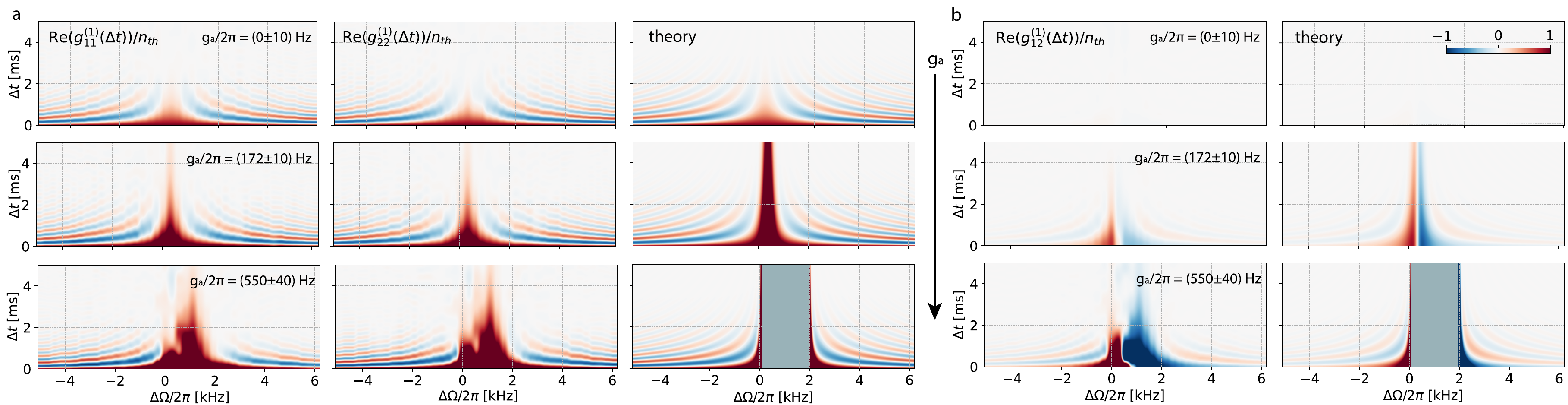}
    \caption{\textbf{First-order correlations: comparison between the experiment and theory.} Correlation functions are plotted as a function of the mechanical detuning and the time delay for increasing coupling rate $g_a$ from top to bottom. \textbf{a.} The autocorrelations in the left and middle columns correspond to the experimental data from particles 1 and 2, respectively, and the right column corresponds to the theoretical model. \textbf{b.} The experimental and theoretical crosscorrelations are shown in the left and right columns, respectively.} 
    \label{figM:Corr}
\end{figure*}

The time traces of the complex envelopes $a_{1/2}$ were used to calculate the first order correlations $g^{(1)}_{jj}/n_T=\langle a^*_j(t)a_j(t+\Delta t)\rangle/n_T$ for $j\in\{1,2\}$, and $g^{(1)}_{12}/n_T=\langle a^*_2(t)a_1(t+\Delta t)\rangle/n_T$ by using python's "scipy.signal.correlate" function. In Fig.\ref{figM:Corr}a, we show the measured autocorrelation functions of particles 1 and 2 (left and middle column) compared to the theoretical model \eqref{eq:autocorrelation} (right column) for increasing coupling rates $g_a$ from top to bottom.
The autocorrelation is highest at zero time delay, where it corresponds to the motion variance. In the oscillatory phase, it exhibits decaying oscillations as a function of the time delay with a decay rate of half of the damping rate, in excellent agreement with the theory.
The oscillation frequency depends on the intrinsic detuning, therefore for uncoupled particles (upper row), the autocorrelations show an oscillating decay everywhere except at zero detuning. 
The variances increase with the coupling rate in the $\mathcal{PT}$-symmetry-broken phase and eventually lead to the limit cycle phase, where the linearized theory is not valid anymore (gray region in the theory plot, bottom row). Note that the autocorrelations look very similar for both particles, confirming the negligible contribution of the reciprocal coupling component ($g_r\approx 0$), according to the Eq. \eqref{eq:autocorrelation}.
The crosscorrelation color plots as a function of the mechanical detuning and the time delay are compared to the linearized model \eqref{eq:crosscorrelation} for increasing coupling rates from top to bottom in Fig.\ref{figM:Corr}b. The magnitude of the crosscorrelation increases with the coupling rate for both theory and experiment and is maximal at $\Delta t=0$, where it corresponds to the amplitude covariance. Within the $\mathcal{PT}$ symmetry-broken regime, the sign of the correlations changes due to the change in the relative phase delay between the oscillations of the two particles. 

\end{document}